\documentclass[fleqn,10pt]{wlscirep}
\usepackage[utf8]{inputenc}
\usepackage[T1]{fontenc}
\usepackage{multicol}
\newenvironment{Figure}
  {\par\medskip\noindent\minipage{\linewidth}}
  {\endminipage\par\medskip}

\usepackage[font=small,labelfont=bf,
   justification=justified,
   format=plain]{caption}
\usepackage{gensymb}
\usepackage{physics}
\usepackage{mathrsfs}

\title{Feynman's path to Schrödinger (and various other things)}

\author[*]{Bernat Frangi}
\author[*]{Héctor López}

\affil[*]{Authors contributed equally to this work}


\begin{abstract}
    Feynman's path integral formulation arose from his attempt to
    incorporate the Lagrangian framework into quantum mechanics, offering
    what he regarded as a more fundamental perspective than the
    Hamiltonian approach, particularly in the context of quantum
    electrodynamics. Inspired by an analogy proposed by P. A. M. Dirac,
    Feynman showed that Schrödinger's equation could be recovered by
    promoting this analogy to an equality, up to a constant factor. This
    insight laid the foundation for the path integral formalism. In this
    article, we first outline the historical development of dynamical
    frameworks in physics and then trace Feynman's reasoning as he
    constructed his formulation, beginning with the double-slit
    experiment and its interpretation. We subsequently derive the
    Schrödinger equation from the path integral and demonstrate the
    conservation of probability. These final sections aim to provide
    students with a clear link between this elegant yet less commonly
    presented approach and the more standard methods typically taught in
    undergraduate courses.
\end{abstract}

\begin{document}

\flushbottom
\maketitle
\thispagestyle{empty}

\begin{multicols}{2}

	\section*{Introduction}

	This article explores Feynman's path integral formulation of quantum
	mechanics, a powerful alternative to the conventional operator
	evolution approach. We begin by briefly surveying different dynamical
	approaches in physics, contrasting the path integral with classical
	variational principles like Maupertuis' and Hamilton's principles of
	least action.

	Our discussion is anchored by the double-slit experiment, which
	profoundly influenced Feynman's development of these ideas. This
	experiment clearly demonstrates that the total probability of an
	event is not simply the sum of the probabilities of distinct
	alternative paths. To explain the observed wave-like interference
	pattern, Feynman introduced the concept of probability amplitude for
	every possible way an event can occur. His postulates state that the
	total amplitude is the sum of amplitudes for all alternative methods,
	and the observed probability is the absolute square of this total
	amplitude.

	The core of the path integral formulation is that the total amplitude
	(kernel) for a particle's transition between two spacetime points is
	the sum (or integral) of the amplitudes for all possible paths
	connecting them. Each path contributes equally in magnitude, but its
	phase is determined by the classical action along that path, divided
	by the reduced Planck constant, $\hbar$, which is interpreted as the
	quantum of action. In the classical limit, where the action is
	significantly larger than $\hbar$, contributions from paths deviating
	considerably from the classical trajectory cancel due to rapid phase
	oscillations. This elegantly illustrates how classical dynamics
	emerges from this quantum framework.

	We will also delve into the role of the wave function and derive the
	Schrödinger equation directly from the path integral formulation.
	This derivation underscores Feynman's success in bridging his
	Lagrangian-based approach with the established Hamiltonian formalism.

	\section*{Different approaches to dynamics}

	In physics, two different ways of understanding dynamics exist. The
	first and most widely used method in quantum mechanics is
	\textit{operator or state evolution in Hilbert spaces}. We find a
	Hilbert space, define some states and operators, and find the way in
	which these operators or states evolve over time. The second and less
	common method, at least until Feynman introduced his path integral
	formulation, is based on the \textit{variational principle}, which
	consist in finding natural quantities whose extrema\footnote{The real
		paths followed by the system in the configuration space are those for
		which the first order change in these quantities is zero for small
		deviations from such path.} give the laws of motion
	\cite{feynman1964principle}.

	The variational principle first entered classical mechanics in the
	18th century, when French mathematician P. L. M. Maupertuis proposed
	that a physical system follows the path such that that the sum of the
	products of momentum and displacement taken over time was minimized
	\cite{maupertuis1746laws}. That is\footnote{The $\delta$ denotes a
		small variation in the path taken from the initial to the final
		configuration. This notation is used here to signify that $S_0$ is a
		functional, that is, a function of functions (the paths).}:
	\begin{equation}
		\delta S_0 = \delta \int \mathbf{p}\cdot d\mathbf{q} = 0,
	\end{equation}
	where $S_0$ has come to be known as the \textit{abbreviated action}.
	With some simple manipulations, using $d\mathbf{q} = \mathbf{v}\cdot
		dt$, we arrive to:
	\begin{equation}
		S_0 = \int 2T\ dt,
	\end{equation}
	with $T=\mathbf{p\cdot v}/2$ the kinetic energy of the system. So,
	\textit{Maupertuis' principle} is equivalent to:
	\begin{equation}
		\int T\ dt = 0.
	\end{equation}

	This marked an early formulation of a \textit{least action
		principle}\footnote{Let us stress again that, really, \textit{least}
		here means \textit{extremum}.}. Such principles assign a numerical
	value, the \textit{action}, to every possible path between two
	points. The true dynamics of the system under study are then found by
	identifying the path that minimizes this action
	\cite{feynman1964principle}. A significant limitation of this early
	version, however, was its restriction to trajectories of constant
	energy, making it applicable only to systems influenced by
	conservative forces. Following Maupertuis' work, foundational
	advancements in variational calculus by L. Euler and J. L. Lagrange
	paved the way for the definitive principle of least action, which W.
	R. Hamilton introduced in the 19th century. \textit{Hamilton's
		principle}, as it is commonly known, is written in terms of the
	action functional $S$, that now involves the Lagrangian $L$:
	\begin{equation}\label{hamilton-pple}
		\left(\delta S\right)_{\Delta t}=0,\ \text{ where }\
		S\left[\mathbf{q}\right] = \int_{t_1}^{t_2} L\left(\mathbf{\dot
				q}(t),\mathbf{q}(t), t\right)\ dt.
	\end{equation}
	The notation $\Delta t= t_2-t_1$ indicates that we are constraining
	the paths to those that start at time $t_1$ and end at time $t_2$.
	The Lagrangian $L$ is defined as $L = T - V$, where $T$ is the
	kinteic energy of the system and $V$ is its potential energy at each
	point in the path \cite{kibble2004classical}.

	Note that, from (\ref{hamilton-pple}), we can see that Maupertuis'
	principle is just a special case of Hamilton's principle. Taking the
	total energy of the system $E = T + V$ and considering it constant
	(since we have only conservative forces), we have, starting from
	Hamilton's principle:
	\begin{equation}
		\begin{split}
			\delta S &= \delta\int L\ dt = \delta\int \left(T - V\right)\
			dt\\
			&= \delta \int \left(T - E + T\right)\ dt\\
			&= \delta \int 2T\ dt - \cancelto{0}{\delta \int E \ dt} = \delta
			S_0.
		\end{split}
	\end{equation}
	So, considering conservative forces, Hamilton's action reduces to
	Maupertuis' action.

	\section*{The double-slit experiment}

	The purpose of the following sections is to give an overview of
	Feynman's path integral formulation of quantum mechanics, following
	\cite{Feynman1965}. Before proceeding, it will be useful to introduce
	the \textit{double-slit experiment}. This conceptual experiment was a
	primary motivator for Feynman in developing his theory and also
	illuminates the thinking behind its postulates or axioms.

	Consider the setup of \textbf{Figure \ref{fig:d-slit}}. At $A$, we
	have a source of electrons $S$, all of which have the same energy but
	come out in random directions to impinge on a screen $B$. The screen
	$B$ has two holes $1$ and $2$, through which electrons can pass. Some
	distance behind the screen $B$, there is a plane $C$ on which an
	electron detector may be placed at different distances $x$ from the
	center of the screen.

	\begin{Figure}
		\centering
		\vspace{1em}
		\includegraphics[width=0.7\textwidth]{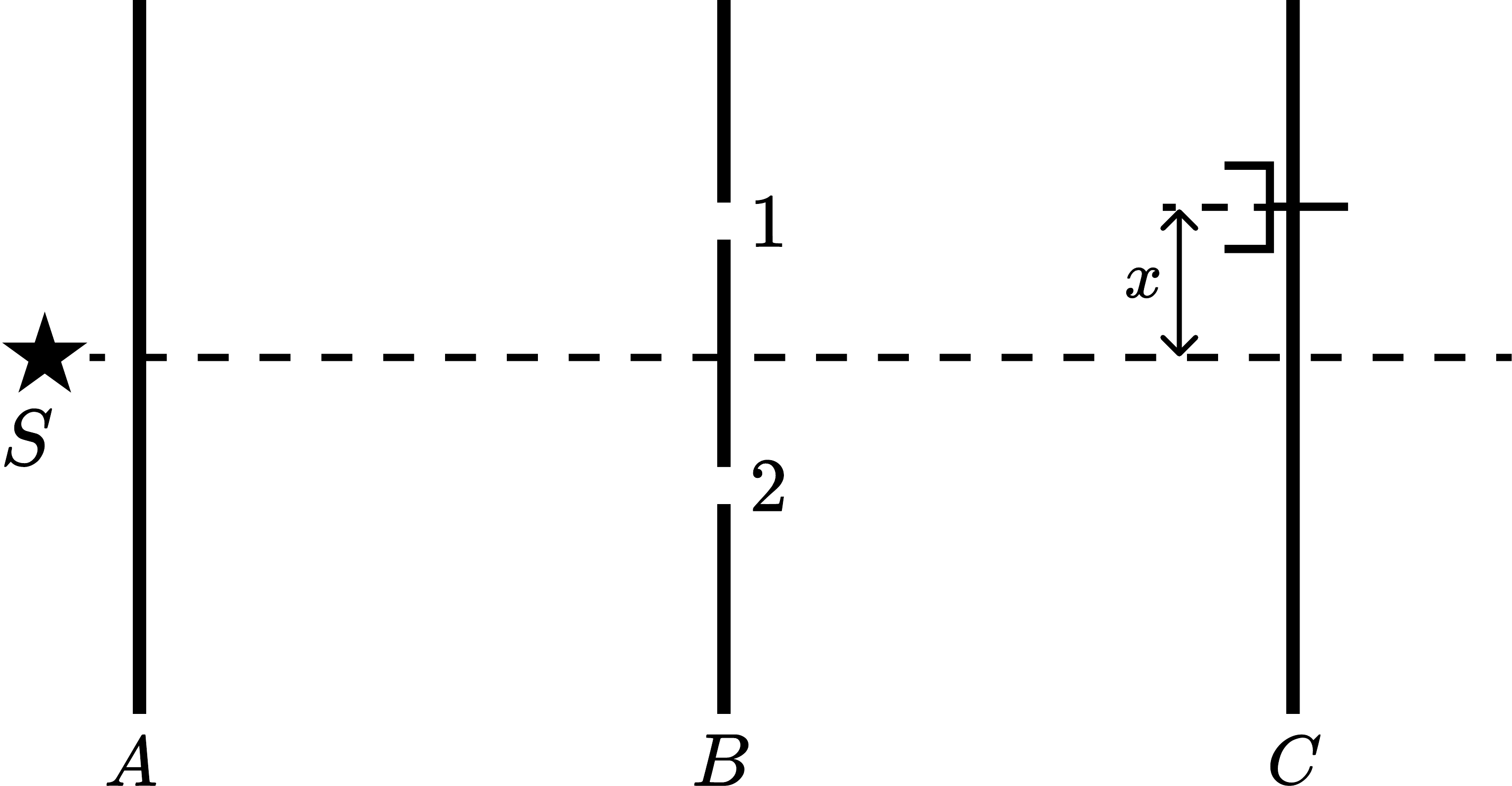}
		\captionof{figure}{Double-slit experiment set-up. Adapted from
			\cite{Feynman1965}.}
		\label{fig:d-slit}
	\end{Figure}

	If the electron source is capable of emitting single electrons and
	the detector is sensitive enough, we find that the current at the
	detector is not continuous, but shows clear peaks corresponding to
	the arrival of each single electron. In between the arrival of each
	electron, nothing will be detected.

	Considering now that we have evenly distributed detectors so that
	they cover the whole of plane $C$, then at the arrival of each
	electron exactly one of the detectors will show a peak in current.
	There will never be half-detections or anything of the sort, which is
	the reason that we consider electrons as particles.

	With these considerations in mind, we can say that the experimental
	set-up of \textbf{Figure \ref{fig:d-slit}} is able to detect the
	passage of single particles (electrons in this case\footnote{Light
		(photons) could be used interchangeably.}) travelling from $S$ to the
	point $x$ through a hole in screen $B$. By moving the detector to
	different values of $x$, we can measure for each position the
	probability $P$ that the electron passes from $S$ to $x$. For
	example, if the source emits one electron per second and we run the
	experiment for $1$ minute, detecting $1$ electron at position $x$ in
	that time interval, the probability $P$ of the electron going from
	$S$ to $x$ will be equal to $1/60$.

	\begin{Figure}
		\centering
		\vspace{1em}
		\includegraphics[width=0.5\textwidth]{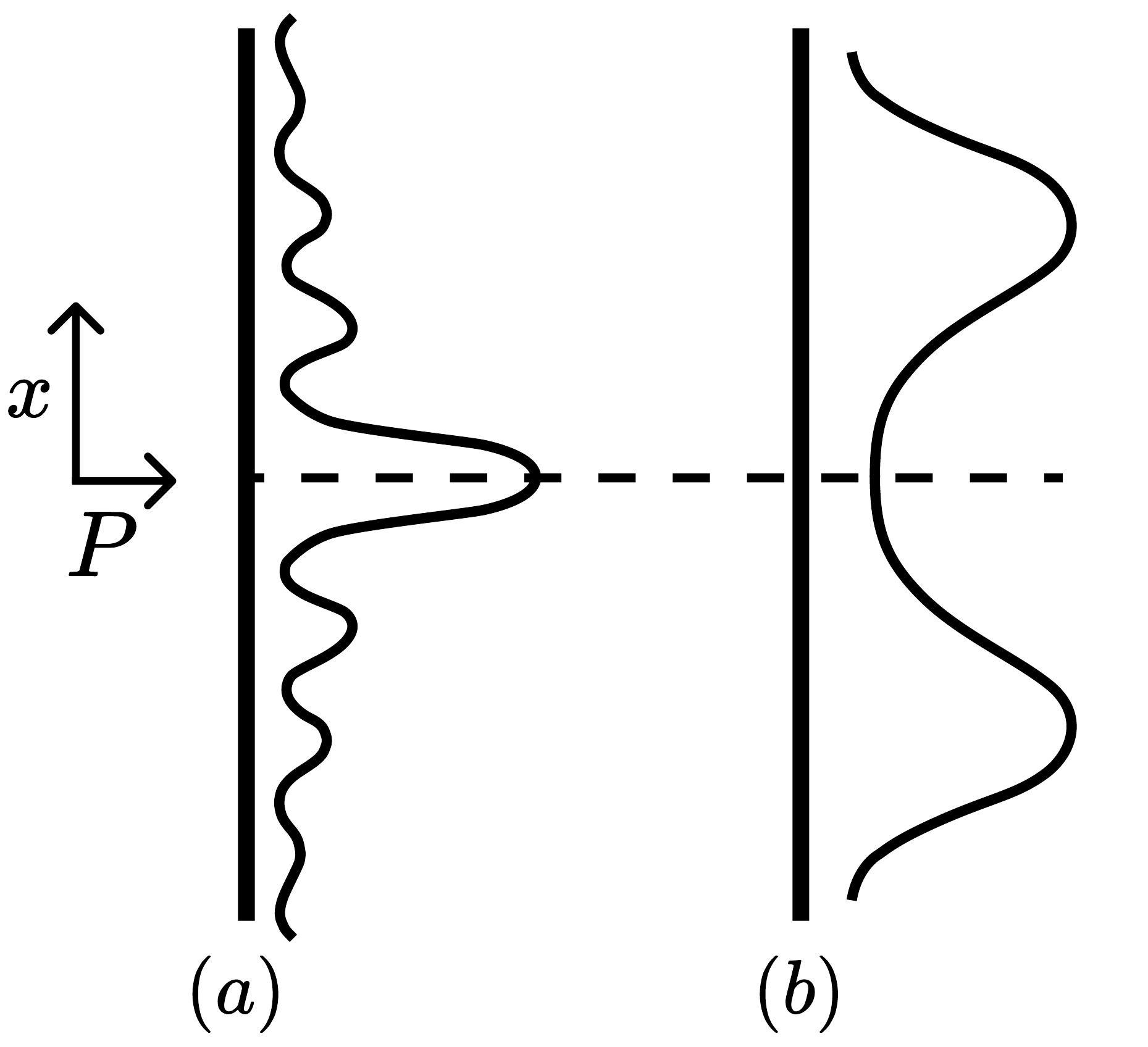}
		\captionof{figure}{Probability $P(x)$ of the electrons going from
			the source $S$ to the screen $C$ as a function of the arriving
			position $x$. (a) shows the real measured probability, while (b)
			shows the probability that we would normally expect in an analogous
			classical experiment (each peak is caused by one of the two holes in
			screen $B$). Adapted from \cite{Feynman1965}.}
		\label{fig:pattern}
	\end{Figure}

	If we now were to plot the probabilities measured as a function of
	$x$, we would be surprised to find the distribution of \textbf{Figure
		\ref{fig:pattern}a}, as opposed to the distribution that we would
	expect from a classical experiment with macroscopic balls, for
	instance, as is shown in \textbf{Figure \ref{fig:pattern}b}.

	What is the meaning of such result? At first, we might suppose the
	following:

	\begin{enumerate}
		\item[(i)] Each electron going from $S$ to $x$ must either do so
			through hole $1$ or through hole $2$. Consequently:
		\item[(ii)] The probability of arriving at $x$ from $S$ should be
			the sum of two parts, each corresponding to the chance of arrival
			through $1$ and through $2$.
	\end{enumerate}

	Therefore, we could repeat the experiment with only hole $1$ open,
	and then with only hole $2$. We could then sum the two distributions
	to obtain the probability of arriving at $x$. However, such a sum
	would give precisely the probability distribution from \textbf{Figure
		\ref{fig:pattern}b}, which is not what we measure when both holes are
	open. Therefore, (ii) must be false. That is, the probability $P$ of
	arriving at $x$ from $S$ is not the sum of the probability of arrival
	through hole $1$ plus that of arrival through hole $2$.

	\section*{Feynman's postulates}

	What is then $P$? To answer this question, we can look again at
	\textbf{Figure \ref{fig:pattern}a} and notice that the pattern is
	precisely the one to be expected if the experiment was to be done
	using waves instead of electrons. By analogy with waves\footnote{This
		analogy is the one that leads us to consider electrons (and photons)
		as waves.}, whose amplitudes are best represented using complex
	numbers, Feynman states in \cite{Feynman1965} the following
	postulates or axioms:

	\vspace{1em}

	\textbf{Postulate 1:} ``There is a quantity called a
	\textit{probability amplitude} associated with every method whereby
	an event in nature can take place''.

	\textbf{Postulate 2:} ``We can associate an amplitude with the
	overall event by adding together the amplitudes of each alternative
	method''.

	\textbf{Postulate 3:} ``The absolute square of the overall amplitude
		[is] the probability that the event will happen''.

	\vspace{1em}

	In other words, for the double-slit experiment, there are two complex
	numbers $\phi_1$ and $\phi_2$ such that
	\begin{equation}\label{pphi_1}
		P =\, \mid \phi_1 + \phi_2 \mid^2,
	\end{equation}
	and the probabilities $P_1$ and $P_2$ of going through hole $1$ and
	hole $2$, respectively, are:
	\begin{equation}\label{pphi_2}
		P_1 = \, \mid \phi_1 \mid^2, \qquad
		P_2 = \, \mid \phi_2 \mid^2.
	\end{equation}

	Thus, if we are able to compute $\phi_1$ and $\phi_2$, we can compute
	the absolute square of $\phi_1+\phi_2$ and interpret it as the
	probability that a particle from $S$ will arrive at $x$. The
	resulting distribution in $x$ is the one in \textbf{Figure
		\ref{fig:pattern}a}.

	\section*{Which hole does the electron go through?}

	Let us now provide some insight into the nature of what we are
	observing in the double-slit experiment.

	According to (\ref{pphi_1}) and (\ref{pphi_2}), it is in general
	\textit{not} true that $P=P_1+P_2$. Then, it is \textit{not} true
	that the particle goes through either hole $1$ or hole $2$ when both
	holes are open. Otherwise, we would be able to classify all electrons
	into two disjoint classes: one for electrons arriving at $x$ through
	hole $1$ and the other for those arriving through hole $2$. Then,
	``the frequency $P$ of arrival at $x$ would surely be the sum of the
	frequency $P_1$ of particles coming through hole $1$ and the
	frequency $P_2$ of those coming through hole $2$''
	\cite{Feynman1965}.

	We could give many possible interpretations for what is actually
	happening with the electron in this experiment. Is the electron going
	through both holes at once? Is it following a complex trajectory
	where it goes first through one hole and then through the other?

	To test these hypotheses, Feynman proposed yet another experiment.
	Since electrons scatter light, we can place a photon source at each
	of the holes $1$ and $2$. In this way, the passage of an electron
	through any one of the holes will cause the scattering of the
	respective photon beam. In doing this experiment, one finds that,
	when an electron passes from $S$ to the screen $C$, light is
	scattered in either hole $1$ or hole $2$, but never in both places.
	It seems that the electron does indeed go through either hole $1$ or
	hole $2$. What is more, if we now use the detectors in screen $C$ to
	find the distribution of $P$ in $x$, we find the curve of
	\textbf{Figure \ref{fig:pattern}b}!

	What is happening? The ``measurement'' of the electron positions at
	the holes using the photon beams is causing the distribution of $P$
	as a function of $x$ to change to $P = P_1+P_2$ (\textbf{Figure
		\ref{fig:pattern}a}), whereas not doing that measurement gives $P\neq
		P_1+P_2$ (\textbf{Figure \ref{fig:pattern}b}). Indeed, the
	interaction of the photons with the electrons causes the chance of
	the electrons arriving at $x$ to change.

	One might think that weakening the photon sources would lead to
	weaker interactions, and thus the change from \textbf{Figure
		\ref{fig:pattern}a} to \textbf{Figure \ref{fig:pattern}b} would be
	somewhat gradual. However, a weaker source would simply mean less
	photons being emitted, and thus some electrons would be missed and
	not scatter any photons. The electrons that did collide would give
	\textbf{Figure \ref{fig:pattern}b}, while those that did not collide
	would give \textbf{Figure \ref{fig:pattern}a}. The resulting pattern
	in screen $C$ would then be some weighed average of both
	distributions. However, each electron would always either fully
	scatter a photon or not at all.

	Another thought would be to use photons of longer wavelengths to
	cause smaller disturbances to the electrons. However, there is always
	a limit to this, since ``a source of light of wavelength $\lambda$
	cannot be located in space with precision greater than $\lambda$. We
	thus see that any physical agency designed to determine through which
	hole the electron passes must produce, lest we have a paradox, enough
	disturbance to alter the distribution'' from \textbf{Figure
		\ref{fig:pattern}a} to \textbf{Figure \ref{fig:pattern}b}
	\cite{Feynman1965}. This idea was first stated by Heisenberg in his
	uncertainty principle.

	In summary of this last discussion, we can add a fourth axiom to
	Feynman's postulates:

	\vspace{1em}

	\textbf{Postulate 4:} ``If we observe the system [...] to be in one
	particular state, we exclude the possibility that it can be in any
	other state, and the amplitudes associated with the excluded states
	can no longer be added in as alternatives in computing the overall
	amplitude'' \cite{Feynman1965}.

	\vspace{1em}

	Furthermore, whether we look at the result of the measurement or not
	is irrelevant. The operation of the measurement equipment is enough
	to disturb the probability amplitude of the system (we could
	obviously look at the result later).

	\section*{The path integral formulation}

	Since the total amplitude is the sum of the amplitude of the
	alternatives, there are many possible ways of evaluating it,
	depending on the different classes we split the alternatives into. In
	one dimension, for instance, we can consider all the alternative
	motions of a particle from a position $x_a$ at time $t_a$ to a
	position $x_b$ at time $t_b$ (we will simply say ``from $a$ to
	$b$''). Each of those alternatives will be given by the position $x$
	as a function of the time $t$, restricted to the conditions $x(t_a) =
		x_a$ and $x(t_b) = x_b$. Each possible \textit{path} would be
	associated to an amplitude, and the sum (or integral) of the
	amplitudes for all possible paths would give the total amplitude of
	the motion.

	Such a sum (or integral) is commonly known as the \textit{kernel} and
	written in \cite{Feynman1965} as $K(b, a)$ for our particular case of
	one dimensional motion.

	Note that, in classical mechanics, there is only a single possible
	trajectory between $a$ and $b$, which we will call the
	\textit{classical trajectory} $\overline x (t)$, following Feynman's
	notation in \cite{Feynman1965}. Such a trajectory is an extremum of
	the action in (\ref{hamilton-pple}), which can be written more simply
	in our one dimensional case as:
	\begin{equation}
		S = \int_{t_a}^{t_b}L(\dot x, x, t)\, dt.
	\end{equation}
	Fixing the endpoints of $x$ and imposing that the first order change
	in $S$ is zero, so $S[x+\delta x] - S[x] = 0$, we arrive at the
	classical Euler-Lagrange equation of motion for one dimension:
	\begin{equation}
		\frac{d}{dt}\left(\frac{\partial L}{\partial \dot x}\right) -
		\frac{\partial L}{\partial x} = 0,
	\end{equation}
	which gives the \textit{unique} path followed by a classical particle
	with Lagrangian $L$.

	The key difference in quantum mechanics is that it is not just this
	path of extreme action that contributes to the amplitude, but
	actually \textit{all} the paths contribute to the total amplitude, or
	kernel, $K(b, a)$. Their contribution is equal in magnitude, but
	different in phase. The phase contribution for a particular path is
	given by the action $S$ for that particular path divided by $\hbar$,
	which we can now interpret as the \textit{quantum of action}, with
	units $J\cdot s$. Thus, we can write the last postulate of Feynman's
	theory, also in his own words from \cite{Feynman1965}:

	\vspace{1em}

	\textbf{Postulate 5:} The different alternatives ``contribute equal
	amounts to the total amplitude, but contribute at different phases.
	The phase of the contribution from a given path is the action $S$ for
	that path in units of the quantum of action $\hbar$''.

	\vspace{1em}

	That is, the probability of going from $a$ to $b$ is given by $P=\,
		\mid K(b, a) \mid^2$, where the kernel $K(b, a)$ is the sum of the
	amplitudes corresponding to all paths $x(t)$ from $a$ to $b$:
	\begin{equation}\label{sum-paths}
		K(b, a) = \sum_{\genfrac{}{}{0pt}{}{\text{all paths
					$x(t)$}}{\text{from $a$ to $b$}}} \phi\left[x(t)\right],
	\end{equation}
	and:
	\begin{equation}
		\phi \left[x(t)\right] = c\cdot \exp\left(i\
		\frac{S\left[x(t)\right]}{\hbar}\right), \qquad c\in \mathbb{C}.
	\end{equation}
	The constant $c$ is a normalization factor that needs to be carefully
	chosen.

	Note that the classical limit corresponds to systems with large
	dimensions, masses, times, etc., so that $S$ is also very large in
	comparison to $\hbar$. In other words, classical systems are those
	involving huge numbers of action quanta, meaning that $S/\hbar$ is a
	very large phase.

	Small deviations $\delta x$ in the path will cause small changes in
	$S$ at the classical scale. However, these changes will be huge in
	the scale of $\hbar$. The rapid oscillations associated with $\delta
		x$ imply that when one path contributes positively, an
	infinitesimally close path will contribute equally but with the
	opposite sign.

	For these cancellations not to occur, $S$ must remain constant (or
	change slowly) with respect to $\delta x$. But this is precisely the
	condition that defines $\overline x$, since it is an extremum of $S$.
	Therefore, the only place where the variation of $S$ with $x$ is
	zero, at least in the first order, is in the neighbourhood of
	$\overline{x}$, where $S\left[\overline x+\delta x\right] =
		S\left[\overline{x}\right]$.

	Thus, for classical systems, the only path that really contributes to
	the amplitude of a certain motion is the classical path
	$\overline{x}$. In this way, the classical laws of motion arise from
	the quantum laws.

	\section*{The path integral}

	The sum over all paths in (\ref{sum-paths}) has been treated
	qualitatively up until now, with the actual mathematical details
	thrown under the carpet\footnote{Note, for instance, that the number
		of paths is a high order of infinity.}. Now, a more precise
	definition of ``sum over all paths'' is required.

	For this, the reader is encouraged to take a look at \textbf{Section
		2.4} of \cite{Feynman1965}. Here, the key idea is to divide the time
	interval $t_b-t_a$ into $N$ infinitesimal time intervals of length
	$\epsilon$, giving a set of time values $t_i$ spaced a distance
	$\epsilon$ apart. To each of these time values $t_i$, a position
	$x_i$ can be assigned. A path then consists of joining the points
	$x_i$ for $i$ between $1$ and $N-1$. The end points $x_0=x_a$,
	$x_N=x_b$, $t_0=t_a$ and $t_N=t_b$ are fixed and $N\epsilon =
		t_b-t_a$. The sum over all paths now consists of integrating over all
	possible choices of the positions $x_i$ and taking the limit $N\to
		\infty$.

	The notation used by Feynman to indicate the path integral from
	(\ref{sum-paths}) is:
	\begin{equation}
		K(b,a) = \int_a^b e^{i S[b,a]/\hbar} \, \mathscr{D}x(t),
	\end{equation}
	where $S[b, a]$ denotes the action of each different path $x(t)$ from
	$a$ to $b$ as we integrate. This expression needs to be properly
	normalized in each case.

	For any path $x(t)$ between $a$ and $b$, we may choose $t_c \in (t_a,
		t_b)$. Then, the action along the path $x(t)$ between $a$ and $b$
	going through $x_c = x(t_c)$ at time $t_c$ can be expressed as
	\begin{equation}
		S[b,a]=S[b,c]+S[c,a].
	\end{equation}
	Then we can rewrite the kernel as
	\begin{equation}
		K(b,a)=\int\exp{\frac{i}{\hbar}(S[b,c]+S[c,a])}\mathcal{D}[x(t)],
	\end{equation}
	where $\mathcal{D}x(t)$ now indicates integration over all paths
	going from $a$ to $c$, integration over all paths going from $c$ to
	$b$, and integration over all possible intermediate positions $x_c$.
	This can be rewritten as:
	\begin{equation}
		K(b,a)=\int_{x_c} K(b,c)K(c,a)dx_c.
	\end{equation}
	A simple extension can be made for an arbitrary number of
	intermediate positions. Considering a path from $a$ to $b$, split
	into $N$ path segments with intermediate positions $x(t_i)=x_i$ for
	$i=1, 2, ..., N-1$, we obtain:
	\begin{equation}
		\begin{split}
			K(b, a) = \int_{x_1}& \int_{x_2} \cdots \int_{x_{N-1}}\, K(b, N-1)\,
			K(N-1, N-2)\, \cdots \\
			& \cdots\, K(i+1, i)\, \cdots\, K(1, a)\, dx_1\, dx_2 \cdots
			dx_{N-1}.
		\end{split}
	\end{equation}
	Thus, we reach the following rule:

	\vspace{1em}

	\textbf{Rule 1:} ``Amplitudes for events occurring in succession in
	time multiply'' \cite{Feynman1965}.

	\vspace{1em}

	If we split the time interval $[t_a, t_b]$ into infinitesimally
	separated time points, with $\Delta t = t_{i+1}-t_i \to \epsilon$,
	then, for sufficiently small $\epsilon$, it holds:
	\begin{equation}
		\begin{split}
			S[i+1, i] &= \int_{t_i}^{t_{i+1}} L\left(\dot x, x, t\right)\
			dt\\
			&=\epsilon \cdot L\left(\frac{x_{i+1}-x_i}{\epsilon},
			\frac{x_{i+1} + x_i}{2}, \frac{t_{i+1} + t_i}{2} \right),
		\end{split}
	\end{equation}
	which is just the length of the time interval multiplied by the
	average value of the Lagrangian in that interval, and is correct to
	the first order in $\epsilon$. With this, we can write the kernel for
	each infinitesimal interval as
	\begin{equation}
		K(i+1, i) =\, \frac{1}{A} \exp \left[ \frac{i\epsilon}{\hbar}\cdot
			L\left( \frac{x_{i+1} - x_i}{\epsilon}, \frac{x_{i+1} + x_i}{2},
			\frac{t_{i+1} + t_i}{2} \right) \right],
		\label{approx}
	\end{equation}
	which is correct to first order in $\epsilon$. The $1/A$ is just a
	normalization constant.
	Applying \textbf{Rule 1}, we obtain:
	\begin{equation}
		\phi[x(t)] = \lim_{\epsilon \to 0} \prod_{i=0}^{N-1} K(i+1, i),
	\end{equation}
	which is the amplitude of a complete path.

	\section*{The wave function}

	In previous sections, we have derived the amplitude for a certain
	motion from a point $a$ to a point $b$. However, it is often useful
	to define this amplitude without consideration of the motion required
	in getting to the final destination $(x,t)$. Let this amplitude be
	$\psi(x,t)$, and the corresponding probability $|\psi(x,t)|^2$. We
	may also call it a \textit{wave function}. The difference to the
	\textit{other} amplitudes is simply a matter of notation reinforcing
	the idea that in this case we are not interested in the previous
	motion of the particle.

	In fact, the kernel $K(b, a) $ is also wave function. While it is
	true that it includes some extra information (namely, that the
	particle comes from $(x_a, t_a)$), it is exactly the amplitude to get
	to $(x_b,t_b)$.

	Since the wave function is an amplitude, it must satisfy the rules
	for combination of amplitudes for events occurring in succession.
	This means it must satisfy the equation
	\begin{equation}
		\psi(x_b,t_b)=\int_{x_c} K(b, c)\psi(x_c,t_c)\, dx_c.
		\label{eq-wf}
	\end{equation}
	That is, the total amplitude to arrive at $(x_b,t_b)$ is the sum,
	over all possible values of $x_c$, of the total amplitude to arrive
	at the point $(x_c,t_c)$ multiplied by the amplitude of going from
	$c$ to $b$. This allows us to express the effects of all the past
	history of the particle in a single function $\psi$. One need only
	know $\psi$ at a particular time to be able to compute the amplitude
	of arriving at any future point.

	\section*{Schrödinger's equation}

	In practice, there are many cases in which evaluating the path
	integral is very difficult, and not at all practical. For those
	cases, there exists the possibility of reducing the path integrals to
	differential equations. In fact, the Hamiltonian formalism of quantum
	mechanics works precisely with one such differential equation called
	the Schrödinger equation. The triumph of Feynman was in his ability
	to show that the Schrödinger equation could be derived from his path
	integral formulation.

	In what follows, we will replicate this derivation, building on what
	we already know about Feynman's path integrals.

	Let us consider two times $t_1=t$ and $t_2=t+\epsilon$ which are
	infinitesimally separated, $t_2-t_1 = \epsilon$. Consider the
	particle is at $y$ at $t_1$ and at $x$ at $t_2$. We can write
	(\ref{eq-wf}) using (\ref{approx}) as
	\begin{equation}\label{schr-1}
		\psi(x,\, t + \epsilon)=\frac{1}{A} \int_{\mathbb{R}}
		\exp\hspace{-0.05cm} \left[ \frac{i \epsilon}{\hbar}\
			L\hspace{-0.05cm}\left( \frac{x - y}{\epsilon}, \frac{x + y}{2} , \,
			t+\cancel{\frac{\epsilon}{2}}\right)
			\hspace{-0.05cm}\right]\hspace{-0.05cm} \psi(y,\, t)\, dy.
	\end{equation}
	We consider now the particular case of a particle in one dimension
	moving in a potential $V(x,\, t)$. The Lagrangian is then:
	\begin{equation}
		L=\frac{1}{2}\, m\, \dot{x}\, ^2-V(x,t).
	\end{equation}
	Substituting in (\ref{schr-1}), we find:
	\begin{equation}
		\begin{aligned}
			\psi(x,\, t + \epsilon) = \frac{1}{A}\int_\mathbb{R} &
			\left\{ \exp\left[ \frac{i m}{\hbar}\frac{(x - y)^2}{2 \epsilon}
			\right] \right\}                                                                                                      \\
			                                                     & \left\{ \exp\left[ -\frac{i \epsilon}{\hbar} V\left( \frac{x +
					y}{2},\, t \right) \right] \right\}
			\psi(y, t)\, dy.
		\end{aligned}
	\end{equation}
	Having $(x-y)^2/\epsilon$ in the first term, we will only find
	relevant contributions of this integral when $y\to x$. This is
	because, as we discussed for the classical limit of the path
	integral, large values of $x-y$ will cause fast oscillations in the
	phase of the exponential term, leading to net cancellations. For this
	reason we may define $y=x+\eta$ with $\eta$ being small enough. Then:
	\begin{multline}
		\psi(x,\, t + \epsilon) =\frac{1}{A} \int_{\mathbb{R}}
		\left\{ \exp\left[ \frac{i m \eta^2}{2 \hbar \epsilon} \right]
		\right\} \\
		\times \left\{ \exp\left[ -\frac{i \epsilon}{\hbar} V\left( \frac{x +
				\eta}{2}, t \right) \right] \right\}
		\psi(x + \eta, t) \, d\eta.
	\end{multline}
	When $\eta$ is of the order $\sqrt{\epsilon\hbar/m}$, the first
	exponential terms changes by the order of $1$ radian. It is then that
	we shall find the relevant contributions to the integral. Also, we
	can let $\epsilon V((x+\eta)/2,t)\to \epsilon V(x,t)$. We now expand
	to first order in $\epsilon$ and second in $\eta$, leaving:
	\begin{multline}
		\psi(x,\, t) + \epsilon\, \frac{\partial \psi}{\partial t} =
		\frac{1}{A}
		\int_{\mathbb{R}}
		e^{\frac{i m \eta^2}{2 \hbar \epsilon}}
		\left[ 1 - \frac{i \epsilon}{\hbar} V(x, t) \right]
		\\ \times \left[ \psi(x, t)
			+ \eta \frac{\partial \psi}{\partial x}
			+ \eta^2 \frac{1}{2} \frac{\partial^2 \psi}{\partial x^2} \right]
		d\eta.
	\end{multline}
	We can separate the right hand side of this equation into three terms
	with the following integrals:
	\begin{equation}
		\frac{1}{A} \int_{\mathbb{R}} e^{i m \eta^2 / 2 \hbar \epsilon} d\eta
		= \frac{1}{A} \left( \frac{2 \pi i \hbar \epsilon}{m} \right)^{1/2},
	\end{equation}
	\begin{equation}
		\frac{1}{A}\int_{\mathbb{R}} e^{i m \eta^2 / 2 \hbar \epsilon} \eta
		\, d\eta = 0,
	\end{equation}
	\begin{equation}
		\frac{1}{A}\int_{\mathbb{R}} e^{i m \eta^2 / 2 \hbar \epsilon} \eta^2
		\, d\eta = \frac{i \hbar \epsilon}{m}.
	\end{equation}
	In order for both sides to agree at the limit $\epsilon \to 0$, we
	must impose $A = ( 2 \pi i \hbar \epsilon/m )^{1/2}$. This is a
	common method of obtaining this constant, also in more complex
	problems. With this, we reach:
	\begin{equation}
		\psi + \epsilon \frac{\partial \psi}{\partial t} = \psi - \frac{i
			\epsilon}{\hbar} V \psi - \frac{\hbar \epsilon}{2 i m}
		\frac{\partial^2 \psi}{\partial x^2}.
	\end{equation}
	Subtracting $\psi$ from both sides:
	\begin{equation}
		\epsilon \frac{\partial \psi}{\partial t} = -\frac{i \epsilon}{\hbar}
		V \psi - \frac{\hbar \epsilon}{2 i m} \frac{\partial^2 \psi}{\partial
			x^2},
	\end{equation}
	and dividing by $\epsilon$:
	\begin{equation}
		\frac{\partial \psi}{\partial t} = -\frac{i}{\hbar} V \psi -
		\frac{\hbar}{2 i m} \frac{\partial^2 \psi}{\partial x^2}.
	\end{equation}
	Finally, multiplying both sides by $-\frac{\hbar}{i}$, we get the
	well-known time-dependent equation for a particle moving in one
	dimension:
	\begin{equation}
		i\hbar\ \frac{\partial \psi(x,\, t)}{\partial t} = -
		\frac{\hbar^2}{2m} \frac{\partial^2 \psi(x,\, t)}{\partial x^2} +
		V(x,\, t) \psi(x,\, t).
	\end{equation}
	In operator form, this is:
	\begin{equation}\label{op-sch-1d}
		i\hbar\ \frac{\partial\psi}{\partial t}=H\psi.
	\end{equation}
	\section*{3D Schrödinger Equation}
	One can generalize the previous procedure using three dimensions to
	get the three dimensional Schrödinger equation. We provide a summary
	of the steps, with the Lagrangian, action and propagator defined in
	three dimensions:
	\begin{equation}
		L(\vec{x},\, \dot{\vec{x}},\, t) = \frac{1}{2}\, m\, \dot{\vec{x}}\,
		^2 - V(\vec{x},\, t),
	\end{equation}
	\begin{equation}
		S[\vec{x}(t)] = \int_{t_a}^{t_b} \left( \frac{1}{2}\, m \,
		\dot{\vec{x}}\, ^2 - V(\vec{x}(t),\, t) \right) dt,
	\end{equation}
	\begin{equation}
		K(\vec x_b,\, t_b;\, \vec x_a,\, t_a) = \int \mathcal{D}[\vec{x}(t)]
		\, e^{\frac{i}{\hbar} S[\vec{x}(t)]}.
	\end{equation}
	Considering that the particle is at $\vec y$ at time $t$ and it is at
	$\vec x $ after an infinitessimal time $\epsilon$, the wave function
	at a time $t +\epsilon$, is:
	\begin{equation}\label{3d-sch-1}
		\psi(\vec{x},\, t + \epsilon) = \int_{\mathbb{R}^3} K(\vec{x},\, t +
		\epsilon;\, \vec{y},\, t) \, \psi(\vec{y}, t) \, d\vec y.
	\end{equation}
	Applying the same approximations used for the one dimensional case:
	\begin{multline}\label{3d-sch-2}
		K(\vec x,\, t + \epsilon;\, \vec y,\, t) \approx
		\\\frac{1}{A} \, \exp \left[ \frac{i \epsilon}{\hbar} \left(
			\frac{m}{2} \left( \frac{\vec{x} - \vec{y}}{\epsilon} \right)^2 -\
			V\left( \frac{\vec{x} + \vec{y}}{2},\, t \right) \right) \right].
	\end{multline}

	We plug (\ref{3d-sch-2}) into (\ref{3d-sch-1}) and, using
	$\vec{y}=\vec{x}+\vec{r}$ with $\mid\mid \vec r\mid\mid $ being small
	enough, we expand $\psi(\vec{y},t)$ around point $\vec x$ and the
	exponential term around $\vec{r}$. We also expand $\psi(\vec x,\,
		t+\epsilon)$ in $\epsilon$. Keeping terms up to second order in $\vec
		r$ and up to first order in $\epsilon$, we get:
	\begin{multline}
		\psi(\vec x,\, t) + \epsilon\, \frac{\partial \psi}{\partial t} =
		\frac{1}{A}
		\int_{\mathbb{R}^3}
		e^{\frac{i m \vec r\, ^2}{2 \hbar \epsilon}}
		\left[ 1 - \frac{i \epsilon}{\hbar} V(\vec x, t) \right]
		\\ \times \left[ \psi(\vec x, t)
			+ \vec r\ \nabla \psi
			+ \vec r\, ^2\, \frac{1}{2}  \nabla^2 \psi \right]
		d\vec r.
	\end{multline}
	Finally, doing similar integrals to the ones in the one dimensional
	case, we obtain the three dimensional Schrödinger equation:
	\begin{equation}
		i \hbar \frac{\partial \psi(\vec{x}, t)}{\partial t} =
		-\frac{\hbar^2}{2m} \nabla^2 \psi(\vec{x}, t) + V(\vec{x}, t)
		\psi(\vec{x}, t).
	\end{equation}

	\section*{Conservation of probability}
	A central requirement of quantum mechanics is the conservation of
	total probability. The integral of the probability density over all
	space must remain constant in time. This ensures physical
	consistency, e.g. that a particle described by the wave function is
	always somewhere in space. In the operator formulation of quantum
	mechanics, this concept follows from the Hermitian nature of the
	Hamiltonian and unitary time evolution. When it comes to path
	integral formulations, conservation of probability manifests in the
	structure of the kernel and how it relates to the probability given
	by the wave function. In this section we examine how the concept of
	conservation of probability arises in Feynman's formalism, and we
	show the necessary condition to be satisfied by the kernel so that
	this is the case.

	The Hamiltonian operator $H$ given by
	\begin{equation}
		H=-\frac{\hbar^2}{2m}\nabla^2+V
	\end{equation}
	is \textit{Hermitian}, as any other Hamiltonian in quantum mechanics.
	Then, it must satisfy (using the one dimensional case)
	\begin{equation}\label{prop-hermit}
		\int (H\psi)^*\psi\ dx=\int \psi ^*(H\psi)\ dx.
	\end{equation}
	As $\psi $ satisfies the Schrödinger equation in (\ref{op-sch-1d}),
	we have:
	\begin{equation}
		\begin{split}
			\int (H\psi)^*\psi\ dx&=\int \left(i\hbar \
			\frac{\partial\psi}{\partial t}\right)^*\psi\ dx\\
			&= -i\hbar \int \frac{\partial\psi^*}{\partial t}\psi\ dx,
		\end{split}
	\end{equation}
	\begin{equation}
		\begin{split}
			\int \psi^*(H\psi)\ dx&=\int \psi^*\left(i\hbar \
			\frac{\partial\psi}{\partial t}\right) dx\\
			&= i\hbar \int\psi^* \frac{\partial\psi}{\partial t}\ dx.
		\end{split}
	\end{equation}
	Combining these results with (\ref{prop-hermit}), we obtain:
	\begin{equation}
		\int \frac{\partial \psi^*}{\partial t} \psi \, dx + \int \psi^*
		\frac{\partial \psi}{\partial t} \, dx = \frac{d}{dt} \left( \int
		\psi^* \psi \, dx \right) = 0.
	\end{equation}
	Then, the ``area under $\psi^*\psi$'', which is just the total
	probability, must be independent of time. That is, the probability is
	constant; it is conserved.

	In what respects to the kernel, this means that, if $f$ is the wave
	function at $t_1$, and $\psi$ is the wave function at $t_2$, so that
	\begin{equation}\label{eqs-two-times}
		\psi(2) = \int K(2,1) f(1) \, dx_1 ,
	\end{equation}
	then the square integral of the wave functions at the different times
	must be the same:
	\begin{equation}\label{sqr-int-wfs}
		\int \psi^*(2) \psi(2) \, dx_2 = \int f^*(1) f(1) \, dx_1.
	\end{equation}
	Substituting (\ref{eqs-two-times}) in (\ref{sqr-int-wfs}), we obtain:
	\begin{multline}
		\iiint K^*(2;\, x_1',t_1) K(2;\, x_1,t_1) f^*(x_1') f(x_1) \,
		dx_1 \, dx_1' \, dx_2 \\
		= \int f^*(x_1) f(x_1) \, dx_1.
	\end{multline}

	Clearly, this equality will hold for any arbitrary $f$ if and only if

	\begin{equation}
		\int K^*(2,x_1',t_1) K(2,x_1,t_1) \, dx_2 = \delta(x_1' - x_1).
	\end{equation}
	Then, if we interpret $\psi$ as a probability amplitude, the kernel
	must satisfy this equation.

	\section*{Conclusion}

	In this work, we have presented the foundations of quantum
	mechanics through Feynman's path integral formalism, tracing his
	reasoning from the double-slit experiment and highlighting key
	insights into his thinking. The article introduces the wave function
	and probability amplitudes, formulates Feynman's postulates, derives
	the Schrödinger equation from the path integral, and examines the
	conservation of probability, linking norm preservation to the
	structure of the propagator. We emphasize how this formalism offers
	intuitive explanations for interference and other quantum phenomena
	and show its connection to the Hamiltonian approach. In doing so, the
	article serves as a pedagogical bridge, helping learners appreciate
	the equivalence of the two formalisms, deepen their conceptual
	understanding with a unifying perspective, and prepare for more
	advanced topics in quantum theory.

\end{multicols}

\bibliography{main}

\section*{Acknowledgements}

The authors would like to thank Prof. Alberto Ibort for his valuable 
insights and stimulating discussions.

\end{document}